\documentclass[preprint2]{aastex62}
\usepackage{color}
\usepackage[titletoc]{appendix}
\usepackage[fleqn]{amsmath}
\usepackage{multirow}

\newcommand{\unit}[1]{\ensuremath{\, \mathrm{#1}}}
\newcommand\tess{\textit{TESS}}
\newcommand\kep{\textit{Kepler}}
\newcommand\gaia{\textit{Gaia}}
\newcommand\jwst{\textit{JWST}}

\shortauthors{Ca\~nas et al.}
\shorttitle{TOI-150: A transiting hot Jupiter in the \tess{} SCVZ}

\begin{document} 

\title{TOI-150: A transiting hot Jupiter in the \tess{} southern CVZ\footnote{This Letter includes data gathered with the 6.5 m \textit{Magellan} Telescopes located at Las Campanas Observatory, Chile.}}

\correspondingauthor{Caleb I. Ca\~nas}
\email{canas@psu.edu}

\author[0000-0003-4835-0619]{Caleb I. Ca\~nas}
\altaffiliation{NASA Earth and Space Science Fellow}
\affiliation{Department of Astronomy \& Astrophysics, The Pennsylvania State University, 525 Davey Lab, University Park, PA 16802, USA}
\affiliation{Center for Exoplanets \& Habitable Worlds, University Park, PA 16802, USA}
\affiliation{Penn State Astrobiology Research Center, University Park, PA 16802, USA}

\author[0000-0001-7409-5688]{Gudmundur Stefansson}
\altaffiliation{NASA Earth and Space Science Fellow}
\affiliation{Department of Astronomy \& Astrophysics, The Pennsylvania State University, 525 Davey Lab, University Park, PA 16802, USA}
\affiliation{Center for Exoplanets \& Habitable Worlds, University Park, PA 16802, USA}
\affiliation{Penn State Astrobiology Research Center, University Park, PA 16802, USA}

\author{Andrew J. Monson}
\affiliation{Department of Astronomy \& Astrophysics, The Pennsylvania State University, 525 Davey Lab, University Park, PA 16802, USA}
\affiliation{Center for Exoplanets \& Habitable Worlds, University Park, PA 16802, USA}

\author{Johanna K. Teske}
\altaffiliation{Hubble Fellow}
\affiliation{Observatories of the Carnegie Institution for Science, 813 Santa Barbara Street, Pasadena, CA 91101, USA}

\author[0000-0003-4384-7220]{Chad F. Bender}
\affiliation{Department of Astronomy and Steward Observatory, University of Arizona, Tucson, AZ 85721, USA}

\author[0000-0001-9596-7983]{Suvrath Mahadevan}
\affiliation{Department of Astronomy \& Astrophysics, The Pennsylvania State University, 525 Davey Lab, University Park, PA 16802, USA}
\affiliation{Center for Exoplanets \& Habitable Worlds, University Park, PA 16802, USA}
\affiliation{Penn State Astrobiology Research Center, University Park, PA 16802, USA}

%Alphabetical
\author[0000-0003-1822-7126]{Conny Aerts}
\affiliation{Instituut voor Sterrenkunde, KU Leuven, Celestijnenlaan 200D, B-3001 Leuven, Belgium}

\author[0000-0002-1691-8217]{Rachael L. Beaton}
\altaffiliation{Hubble Fellow}
\altaffiliation{Carnegie-Princeton Fellow}
\affiliation{Department of Astrophysical Sciences, 4 Ivy Lane, Princeton University, Princeton, NJ 08544, USA}
\affiliation{Observatories of the Carnegie Institution for Science, 813 Santa Barbara Street, Pasadena, CA 91101, USA}

\author[0000-0003-1305-3761]{R. Paul Butler}
\affiliation{Department of Terrestrial Magnetism, Carnegie Institution for Science, 5241 Broad Branch Road, NW, Washington, DC 20015, USA}

\author[0000-0001-6914-7797]{Kevin R. Covey}
\affil{Department of Physics \& Astronomy, Western Washington University, Bellingham, WA 98225, USA}

\author[0000-0002-5226-787X]{Jeffrey D. Crane}
\affiliation{Observatories of the Carnegie Institution for Science, 813 Santa Barbara Street, Pasadena, CA 91101, USA}

\author[0000-0002-3657-0705]{Nathan De Lee}
\affiliation{Department of Physics, Geology, and Engineering Technology, Northern Kentucky University, Highland Heights, KY 41099, USA}
\affiliation{Department of Physics \& Astronomy, Vanderbilt University, Nashville, TN 37235, USA}

\author[0000-0002-2100-3257]{Mat\'ias R. D\'iaz}
\affiliation{Observatories of the Carnegie Institution for Science, 813 Santa Barbara Street, Pasadena, CA 91101, USA}
\affiliation{Universidad de Chile, Departmento de Astronom\'ia,
Camino El Observatorio 1515, Las Condes, Santiago, Chile}

\author[0000-0003-0556-027X]{Scott W. Fleming}
\affiliation{Space Telescope Science Institute, 3700 San Martin Drive, Baltimore, MD 21218, USA}

\author[0000-0002-1693-2721]{D. A. Garc\'ia-Hern\'andez}
\affiliation{Instituto de Astrof\'isica de Canarias (IAC), E-38205 La Laguna, Tenerife, Spain}
\affiliation{Universidad de La Laguna (ULL), Departamento de Astrof\'isica, E-38206 La Laguna, Tenerife, Spain}

\author[0000-0002-1664-3102]{Fred R. Hearty}
\affiliation{Department of Astronomy \& Astrophysics, The Pennsylvania State University, 525 Davey Lab, University Park, PA 16802, USA}
\affiliation{Center for Exoplanets \& Habitable Worlds, University Park, PA 16802, USA}

\author{Juna A. Kollmeier}
\affiliation{Observatories of the Carnegie Institution for Science, 813 Santa Barbara Street, Pasadena, CA 91101, USA}

\author[0000-0003-2025-3147]{Steven R. Majewski}
\affiliation{Department of Astronomy, University of Virginia, Charlottesville, VA 22904, USA}

\author{Christian Nitschelm}
\affiliation{Centro de Astronom{\'i}a (CITEVA), Universidad de Antofagasta, Avenida Angamos 601, Antofagasta 1270300, Chile}

\author{Donald P. Schneider}
\affiliation{Department of Astronomy \& Astrophysics, The Pennsylvania State University, 525 Davey Lab, University Park, PA 16802, USA}
\affiliation{Center for Exoplanets \& Habitable Worlds, University Park, PA 16802, USA}

\author{Stephen A. Shectman}
\affiliation{Observatories of the Carnegie Institution for Science, 813 Santa Barbara Street, Pasadena, CA 91101, USA}

\author[0000-0002-3481-9052]{Keivan G. Stassun}
\affiliation{Department of Physics \& Astronomy, Vanderbilt University, Nashville, TN 37235, USA}

\author[0000-0003-0842-2374]{Andrew Tkachenko}
\affiliation{Instituut voor Sterrenkunde, KU Leuven, Celestijnenlaan 200D, B-3001 Leuven, Belgium}

\author[0000-0002-6937-9034]{Sharon X. Wang}
\affiliation{Department of Terrestrial Magnetism, Carnegie Institution for Science, 5241 Broad Branch Road, NW, Washington, DC 20015, USA}

\author{Songhu Wang}
\altaffiliation{\textit{51 Pegasi b} Fellow}
\affiliation{Department of Astronomy, Yale University, New Haven, CT 06511, USA}

\author{John C. Wilson}
\affiliation{Department of Astronomy, University of Virginia, Charlottesville, VA 22904, USA}

\author{Robert F. Wilson}
\affiliation{Department of Astronomy, University of Virginia, Charlottesville, VA 22904, USA}
\begin{abstract}
We report the detection of a hot Jupiter ($M_{p}=1.75_{-0.17}^{+0.14}\unit{M_{J}}$, $R_{p}=1.38\pm0.04\unit{R_{J}}$) orbiting a middle-aged star ($\log g=4.152^{+0.030}_{-0.043}$) in the \textit{Transiting Exoplanet Survey Satellite} (\textit{TESS}) southern continuous viewing zone ($\beta=-79.59\degr$). We confirm the planetary nature of the candidate TOI-150.01 using radial velocity observations from the APOGEE-2 South spectrograph and the Carnegie Planet Finder Spectrograph, ground-based photometric observations from the robotic Three-hundred MilliMeter Telescope at Las Campanas Observatory, and \textit{Gaia} distance estimates. Large-scale spectroscopic surveys, such as APOGEE/APOGEE-2, now have sufficient radial velocity precision to directly confirm the signature of giant exoplanets, making such data sets valuable tools in the \textit{TESS} era. Continual monitoring of TOI-150 by \textit{TESS} can reveal additional planets and subsequent observations can provide insights into planetary system architectures involving a hot Jupiter around a star about halfway through its main-sequence life.
\end{abstract}
\keywords{planetary systems --- techniques: spectroscopic --- techniques: photometric}
\section{Introduction}
The \textit{Transiting Exoplanet Survey Satellite} \citep[\tess{},][]{Ricker2015} is an ongoing mission designed to survey the entire sky and discover transiting exoplanets around bright, nearby stars. These exoplanets are ideal targets for high-precision spectroscopic observations to provide mass estimates, and for future space- and ground-based atmospheric characterization. Approximately 1000 planets are expected to be detected in the \tess{} full-frame images around relatively bright stars with \tess{} magnitudes ($T$) $\sim11$ \citep[e.g.,][]{Barclay2018}. \tess{} divides the sky into 26 sectors rotating about the ecliptic poles and will observe stars within \(\sim12\degr\) of the poles (the continuous viewing zone, CVZ) for \(\sim351\) days. 

In this Letter, we confirm the planetary nature of the candidate TOI-150.01 (TIC 271893367, 2MASS J07315176-7336220, \gaia{} DR2 5262709709389254528, $T=10.87$, $V=11.39$, $H=10.05$) using \tess{} photometry, ground-based photometric observations, \gaia{} DR2 distance estimates, and Doppler velocimetry from the Sloan Digital Sky Survey (SDSS)/APOGEE-2 South spectrograph and the Carnegie Planet Finder Spectrograph (PFS). This Letter is structured as follows. Section \ref{sec:observations} presents the observational data and data processing. Section \ref{sec:fpp} discusses stellar contamination and constraints on binarity and stellar companions, and Section \ref{sec:models} describes our analysis and derivation of the system parameters. A discussion of our results is presented in Section \ref{sec:discussion}.
\section{Observations and Data Reduction}\label{sec:observations}
\subsection{TESS Photometry}
TOI-150 was observed by \tess{} in Sectors \(1-4\) and has one planetary candidate, TOI-150.01, with a period of \(\sim5.85\) days derived from the ``quick-look pipeline'' developed by the MIT branch of the \tess{} Science Office (C. X. Huang et al. 2019, in preparation). The \tess{} full-frame image light curves were corrected for systematic trends using the difference imaging analysis toolset described by \cite{Oelkers2018}. The corrected and extracted light curves are available\footnote{\url{https://filtergraph.com/tess_ffi}} through the {\tt Filtergraph} data visualization service \citep{Burger2013}, where the flux for each star is extracted using a fixed aperture radius of 3.5 pixels (\(\sim73.5''\)) and sky annulus radii of 5-7 pixels (\(105''-147''\)). We use the ``clean'' photometry from \cite{Oelkers2019}, where the light curve has been corrected for systematics using the median trend apparent in 100 other stars of comparable magnitude and located at least 10 \tess{} pixels (\(\sim210''\)) from the star. \cite{Oelkers2019} warned that some residual variability common to those 100 stars may be injected into a target's ``clean'' light curve. The ``clean'' light curve for TOI-150 exhibited some residual variability and was detrended using a Gaussian process as described in \cite{Canas2019}. There was no further processing of the photometry and no Gaussian process was employed when fitting the photometry and velocimetry.
\subsection{TMMT Ground-based Photometry}
We observed two transits of TOI-150.01 using the robotic Three-hundred MilliMeter Telescope \citep[TMMT;][]{Monson2017} at Las Campanas Observatory (LCO) on the nights of 2018 December 13 and 20. Both observations were performed slightly out of focus in the Cousins $I_{C}$ filter \citep {Doi2010}, resulting in a point spread function FWHM of $4\arcsec$. We obtained 221 and 200 frames for the December 13 and 20 observations, respectively, using an exposure time of 120s. In the 1x1 binning mode used, the detector has a 13s readout time between exposures, resulting in an effective cadence of 133s and an observing efficiency of 90\%. The December 13 observations began at airmass 2.18, reached 1.40 at the meridian, and ended at airmass 1.48. For the December 20 observations, the corresponding airmasses are 1.96, 1.40, and 1.48.

We processed the photometry using AstroImageJ \citep{Collins2017} following the procedures described in \cite{Stefansson2017}. After experimenting with a number of different apertures, we adopted an object aperture radius of 10 pixels (11\arcsec), and inner and outer sky annulii of 14 pixels (17\arcsec) and 21 pixels (25\arcsec), respectively, for both nights. These values minimized the standard deviation in the residuals for each observation. Following \cite{Stefansson2017}, we added the expected scintillation-noise errors to the photometric error (including photon, readout, dark, sky background, and digitization noise).
\subsection{Spectroscopic Observations}
\label{radsec}
TOI-150 was observed from the Carnegie Observatory's LCO on 2018 January 28 and 30 as part of a systematic survey of the \tess{} southern CVZ carried out using the southern spectrograph of the APO Galaxy Evolution Experiment \citep[APOGEE;][]{Majewski2017,Zasowski2017}. This survey was initiated as an APOGEE-2S external program led by Carnegie Observatories, which contributed data beyond the scope of the original galactic evolution goals of the SDSS-IV southern survey \citep{Blanton2017}. Both spectra of TOI-150 were obtained with the high-resolution (\(R=\Delta\lambda/\lambda\sim22,500\)), near-infrared (\(1.51-1.7\) \unit{\mu m}), multi-object APOGEE-2S spectrograph \citep{Wilson2019} mounted on the Ir\'en\'ee du Pont 2.5 m telescope \citep{Bowen1973}. For each observation, the APOGEE data  pipeline \citep{Nidever2015} performs sky subtraction, telluric and barycentric correction, and wavelength and flux calibration. The radial velocities (RVs) were derived using a maximum-likelihood cross-correlation method with BT-Settl synthetic spectra \citep{Allard2012} following the procedure described in \cite{Canas2019}. 

\startlongtable
\begin{deluxetable*}{ccccc}
\tablecaption{Spectroscopic Observations\label{tab:table1}}
\tablehead{\colhead{\multirow{ 2}{*}{BJD\(_\text{TDB}^{a}\)}} & 
\colhead{RV} & 
\colhead{\(1\sigma\)} & 
\colhead{S/N\(^b\)}\\
\colhead{} & 
\colhead{(m s\(^{-1}\))} & 
\colhead{(m s\(^{-1}\))} & 
\colhead{(pixel\(^{-1}\))}
}
\startdata
\sidehead{APOGEE-2S:}
~~2458146.705799 & 4666 & 77 & 168 \\
~~2458148.693524 & 5073 & 80 & 146  \\
\sidehead{PFS$^{c}$:}
~~2458476.801377 & 151.15 & 2.27 & 52 \\
~~2458479.785058 & $-176.59$ & 2.25 & 49 \\
~~2458501.708368 & 25.43 & 2.46 & 43 \\
\enddata
\tablenotetext{a}{BJD\(_\text{TDB}\) is the Barycentric Julian Date in the Barycentric Dynamical Time standard}
\tablenotetext{b}{APOGEE-2S has \(\sim2\) pixels per resolution element.}
\tablenotetext{c}{PFS velocities are relative; the uncertainties are the internal errors from the PFS pipeline.}
\end{deluxetable*}
\section{Stellar Companions to TOI-150}\label{sec:fpp}
\subsection{Sky-projected Stellar Companions}
\begin{figure*}[!ht]
\epsscale{1.1}
\plotone{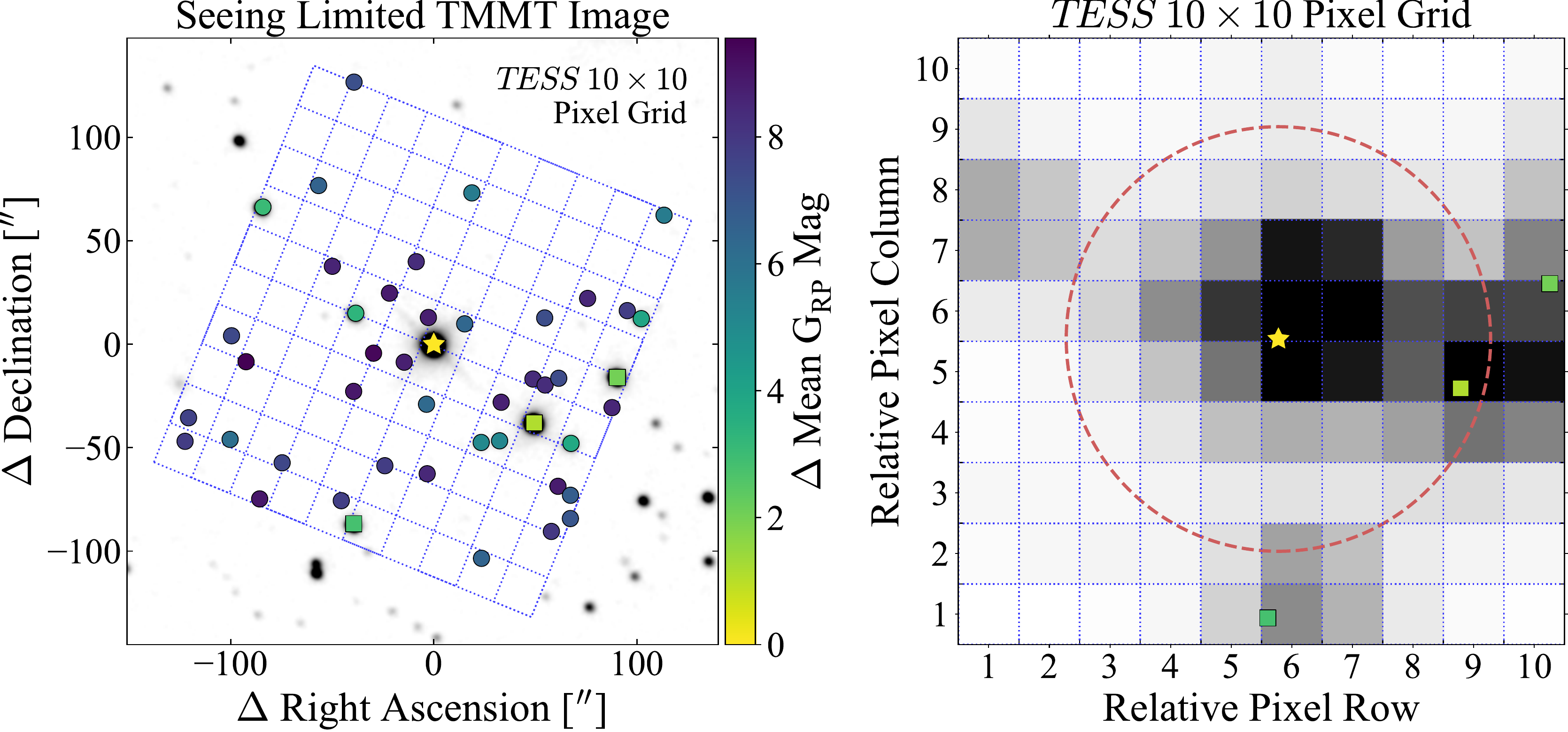}
\caption{Stellar neighborhood of TOI-150. The left panel shows the 47 other stars identified in \gaia{} DR2 inside the $10\times10$ \tess{} pixel grid from Sector 1 plotted above our seeing-limited image from TMMT (TOI-150 is denoted by the star). The right panel displays the location of the three stars, shown as squares, that are within three magnitudes of TOI-150 and are found within the $10\times10$ \tess{} grid. The aperture used by \cite{Oelkers2019} to derive the light curve is denoted as a dashed circle. The closest star, TIC 271893376, lies within this aperture and is a source of flux contamination that dilutes the \tess{} light curve. The TMMT aperture does not include this star. \label{fig:f1}}
\end{figure*}

The large aperture radius (\(73.5''\)) used to process the \tess{} photometry ensures there will be flux contamination in the light curves. To investigate the background stars, we used \gaia{} DR2 \citep{GaiaCollaboration2018} and searched a $10\times10$ \tess{} pixel grid centered on TOI-150.  The left panel of Figure \ref{fig:f1} is an in-focus, seeing-limited image from TMMT overlaid with the $10\times10$ \tess{} pixel grid and the stars identified in \gaia{} DR2, each shaded by the difference to the \gaia{} $G_{RP}$ magnitude of TOI-150. The right panel of Figure \ref{fig:f1} is the $10\times10$ \tess{} pixel grid for Sector 1, with all sources within three magnitudes of TOI-150. A total of 47 other stars reside in this region. The brightest stellar neighbor, TIC 271893376 ($T$ = 11.97), is inside the aperture used by \cite{Oelkers2019} at a sky-projected distance of \(62.23''\). TIC 271893376 and all other stars identified by \gaia{} lie outside the aperture used to process the TMMT photometry.
\begin{figure*}[!ht]
\epsscale{1.1}
\plotone{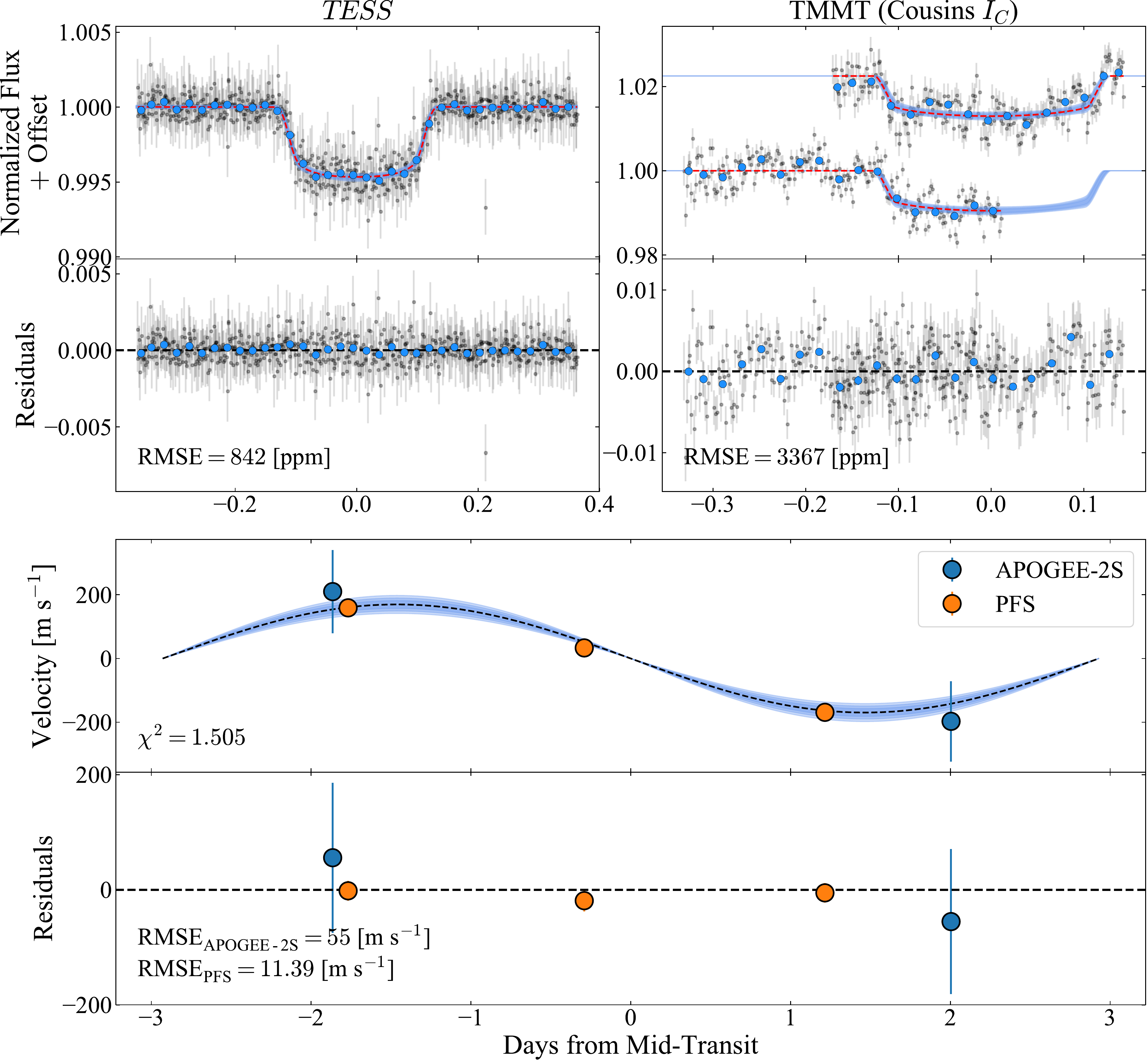}
\caption{Photometry and velocimetry of TOI-150. The top panels display the phased light curves from \tess{} and TMMT with the best models and the root mean square error (RMSE). The small dots are the raw data and the larger circles are the data binned to a 10 minute cadence. An artificial offset is included in the TMMT data to show the two observed transits. The bottom panels show the radial velocities phased to the period of TOI-150.01. The RV RMSEs are formal values that underestimate the true instrumental RMSE for this small data set. The derived systemic velocities have been removed from the data. The uncertainties have been inflated by \(\sim1.64\) and \(\sim7.61\) for APOGEE-2N and PFS, respectively. \label{fig:f2}}
\end{figure*}
\subsection{Non-detection of Spectroscopic Companions within 1.31$''$ of TOI-150}
In lieu of adaptive optics imaging for TOI-150, we use the APOGEE-2S spectra to search for light from secondary stars in the spectrum with the highest signal-to-noise ratio (S/N). The fiber core for APOGEE-2S has a radius of \(1.31''\). We use the software {\tt binspec} \citep{El-Badry2018,El-Badry2018a} to search for the faint spectrum of a second star by modeling the observed stellar spectra as the sum of two input model spectra. The spectrum of the primary star, TOI-150, is fit with a neural network spectral model \citep{Ting2018}. 
The neural network employed by {\tt binspec} was trained on the Kurucz stellar library \citep{Kurucz1979} and is valid in the regime of $4200 \unit{K} < T_{e} < 7000\unit{K}$, $4.0 < \log g < 5.0$, and $−1 < \unit{[Fe/H]} < 0.5$ for slow rotating (\(v_{macro}<45\unit{km\ s^{-1}}\)) main-sequence stars. While {\tt binspec} is designed to fit both single and binary spectra, it is limited to the detection of moderate-mass ratio binaries (\(0.4\lesssim q\lesssim0.85\)). When adopting the model selection statistics and criterion from \cite{El-Badry2018}, there is no indication of a companion, as evidenced by comparison of a single-component fit to a binary component fit, which yields \(\Delta\chi^{2}=9.15\) and \(f_{imp}=2.2\times10^{-5}\). The minimum case for binarity requires \(\Delta\chi^{2}>300\) and \(f_{imp}>0.225\).
\section{System Parameters}\label{sec:models}
\subsection{Stellar Parameters}
The APOGEE Stellar Parameter and Chemical Abundances Pipeline \citep[ASPCAP;][]{GarciaPerez2016} provides spectroscopic stellar parameters for TOI-150. These parameters are derived from a composite APOGEE-2 spectrum, are empirically calibrated, and with the exception of the surface gravity, are determined to be quite reliable \citep[see][]{Holtzman2018}. For TOI-150, ASPCAP provides \(T_{e}=6088\pm130\) K and \(\unit{[Fe/H]}=0.16\pm0.01\). The surface gravity was poorly constrained after calibration and the uncalibrated value is \(\log g=4.47\). 

As the processing of APOGEE-2S data is still in development, we derived an independent set of stellar parameters using the PFS iodine-free ``template''. We employed the {\tt SpecMatch-Emp} algorithm \citep{Yee2017} to characterize the properties of TOI-150 by comparing the optical spectrum to a library of 404 high-resolution (\(R\sim55,000\)), high quality (\(\unit{S/N}>100\)) Keck/HIRES stellar spectra that have well-determined properties. In brief, {\tt SpecMatch-Emp} shifts the observed spectrum to the library wavelength scale, finds the best-matching library spectrum using \(\chi^{2}\) minimization, and uses a linear combination of the five best-matching spectra to synthesize a composite spectrum. Following the examples provided by \cite{Yee2017}, we compared the spectral order containing the Mg triplet line (\(\sim516-520\unit{nm}\)) to the Keck/HIRES library. The derived parameters are \(T_{e}=6029\pm110\) K, \(\log g=4.15\pm0.12\), \(\unit{[Fe/H]}=0.25\pm0.09\). The calibrated ASPCAP values for \(T_{e}\) and [Fe/H] are within the \(1\sigma\) uncertainties from the respective {\tt SpecMatch-Emp} values. To provide a self-consistent set of stellar parameters that include a reliable \(\log g\) value, we adopt the {\tt SpecMatch-Emp} derived parameters for further analysis of TOI-150.
\subsection{System Parameters}
We used the {\tt EXOFASTv2} analysis package \citep{Eastman2017} to model the spectral energy distribution and derive the stellar parameters using MIST stellar models \citep{Choi2016}. We assumed Gaussian priors using the (i) 2MASS \(JHK\) magnitudes \citep{Skrutskie2006}, (ii) SDSS \(g'r'i'\) and Johnson \(BV\) magnitudes from the AAVSO Photometric All-Sky Survey \citep{Henden2015}, (iii) Tycho-2 \(B_{T}V_{T}\) magnitudes \citep{Hog2000}, (iv)  \textit{Wide-field Infrared Survey Explorer} magnitudes \citep{Wright2010}, (v) the host star surface gravity, temperature and metallicity derived with {\tt SpecMatch-Emp}, and (vii) the distance estimate from \cite{Bailer-Jones2018}. We adopt a uniform prior for the maximum visual extinction from estimates of Galactic dust extinction by \cite{Schlafly2011}. The stellar priors and derived stellar parameters with their uncertainties are listed in Table \ref{tab:table3}. The reported values, derived using distance estimates from \cite{Bailer-Jones2018}, are within the \(1\sigma\) uncertainties of the values derived when adopting a Gaussian prior based on the distance from (i) the \gaia{} DR2 parallax, as corrected for a systematic offset, following \cite{Stassun2018a}, or (ii) the Bayesian distance estimate for the \gaia{} RV stellar sample derived by \cite{Schoenrich2019}.

The {\tt juliet} analysis package \citep{Espinoza2018} was employed to jointly model the photometry and velocimetry. {\tt juliet} utilizes publicly available tools to model the photometry \citep[\texttt {batman};][]{Kreidberg2015} and velocimetry \citep[\texttt{radvel};][]{Fulton2018} and performs the parameter estimation using the importance nest-sampling algorithm \texttt {MultiNest} \citep{Feroz2013,Buchner2014}. We validated the performance of our {\tt juliet} implementation by performing an analysis on KOI-189 \citep{Diaz2014}. The resulting parameters were essentially identical to those published by \cite{Diaz2014} using the {\tt PASTIS} planet-validation software. The photometric model is based on the analytical formalism of \cite{Mandel2002} for a planetary transit assuming a quadratic limb-darkening law. The model is modified to include a dilution factor, \(D\), which is the ratio of the out-of-transit flux of the target to that of the total flux from other stars within the photometric aperture. We fit a dilution factor for \tess{} photometry and use the ground-based TMMT observations, where the aperture excludes all detected \gaia{} stellar neighbors, to constrain the true transit depth of TOI-150.01. 

The radial velocity model is a standard Keplerian model. The few spectroscopic observations of TOI-150 cannot constrain eccentricity and we adopt a circular orbit (\(e=0\) and \(\omega=90\degr\)) for TOI-150.01. The uncertainties from Table \ref{tab:table1} are scaled such that the reduced chi-squared statistic, \(\chi^{2}_{\nu}\), for each spectrograph is approximately the 50th percentile of the respective complementary cumulative distribution function. For a circular Keplerian orbit, the only degrees of freedom in our radial velocity model are a fraction of the semi-amplitude and the respective systemic velocity. We took a conservative approach and assumed the APOGEE-2N data only constrained \(10\%\) of the semi-amplitude, \(K\). With these assumptions, the uncertainties listed in Table \ref{tab:table1} were inflated by \(\sim1.64\) and \(\sim7.61\) for APOGEE-2N and PFS, respectively, in the joint fit. We have inflated the radial velocity uncertainties such that the best fit agrees with a circular orbit and we note this overestimates the RV uncertainty if TOI-150.01 were on an eccentric orbit. Figure \ref{fig:f2} presents the result of the fit to the photometry and velocimetry. Table \ref{tab:table3} provides a summary of the stellar priors, together with the inferred system parameters and respective confidence intervals. The uncertainties from the model-dependent stellar parameters were analytically propagated when deriving \(M_{p}\), \(R_{p}\), \(\rho_{p}\), \(T_{eq}\), and \(a\).
\startlongtable
\begin{deluxetable*}{llcc}
\tabletypesize{\normalsize}
\tablecaption{Parameters for the TOI-150 System \label{tab:table3}}
\tablehead{\colhead{~~~Parameter} &
\colhead{Units} &
\multicolumn{2}{c}{Value}
}
\startdata
\sidehead{Stellar Priors:}
~~~Effective Temperature$^{a}$ \dotfill & $T_{e}$ (K)\dotfill & \multicolumn{2}{c}{$6029\pm110$}\\
~~~Surface Gravity$^{a}$ \dotfill & $\log(g)$ (cgs) \dotfill & \multicolumn{2}{c}{$4.15\pm0.12$}\\
~~~Metallicity$^{a}$ \dotfill & [Fe/H] \dotfill & \multicolumn{2}{c}{$0.25\pm0.09$}\\
~~~Maximum Visual Extinction \dotfill & \(A_{V,max}\) \dotfill & \multicolumn{2}{c}{$0.556$}\\
~~~Distance\dotfill & (pc)\dotfill & \multicolumn{2}{c}{$336\pm2$}\\
\sidehead{Derived Model-Dependent Stellar Parameters$^{b}$:}
~~~Mass \dotfill & $M_{*}$ (\unit{M_{\odot}}) \dotfill & \multicolumn{2}{c}{$1.249_{-0.115}^{+0.069}$}\\
~~~Radius\dotfill & $R_{*}$ (\unit{R_{\odot}}) \dotfill & \multicolumn{2}{c}{$1.551_{-0.025}^{+0.024}$}\\
~~~Density \dotfill & $\rho_{*}$ (g \unit{cm^{-3}}) \dotfill & \multicolumn{2}{c}{$0.470_{-0.045}^{+0.040}$}\\
~~~Surface Gravity \dotfill & $\log(g)$ (cgs) \dotfill & \multicolumn{2}{c}{$4.152_{-0.043}^{+0.030}$}\\
~~~Effective Temperature\dotfill & $T_{e}$ (K) \dotfill & \multicolumn{2}{c}{$6003_{-98}^{+104}$}\\
~~~Metallicity\dotfill & [Fe/H] \dotfill & \multicolumn{2}{c}{$0.235_{-0.084}^{+0.083}$}\\
~~~Age\dotfill & (Gyr) \dotfill &
\multicolumn{2}{c}{$4.3_{-1.3}^{+2.5}$}\\
~~~Parallax \dotfill & (mas) \dotfill &
\multicolumn{2}{c}{$2.974\pm0.017$}\\
~~~Visual Extinction \dotfill & $A_{V}$ \dotfill 
& \multicolumn{2}{c}{$0.188_{-0.084}^{+0.082}$} \\
\multicolumn{2}{l}{Derived Photometric Parameters:} & $TESS$ & Cousins $I_{C}$ \\
~~~Linear Limb-darkening Coefficient\dotfill & $u_1$\dotfill & 
$0.21_{-0.14}^{+0.20}$ & $0.41_{-0.20}^{+0.19}$ \\
~~~Quadratic Limb-darkening Coefficient\dotfill & $u_2$\dotfill & 
$0.39_{-0.32}^{+0.27}$ & $-0.05_{-0.17}^{+0.27}$ \\
~~~Dilution Factor\(^{c}\)\dotfill & $D$\dotfill & 
$0.49\pm0.03$ & $\cdots$ \\
\sidehead{Derived Orbital Parameters:}
~~~Orbital Period\dotfill & $P$ (days) \dotfill &
\multicolumn{2}{c}{$5.857342_{-0.000066}^{+0.000065}$}\\
~~~Semi-major Axis\dotfill & $a$ (au) \dotfill &
\multicolumn{2}{c}{$0.0583_{-0.0018}^{+0.0013}$}\\
~~~Semi-amplitude Velocity\dotfill & $K$ (m \unit{s^{-1}})\dotfill &
\multicolumn{2}{c}{$169.58_{-12.38}^{+11.90}$}\\
~~~Mass Ratio\dotfill & $q$ \dotfill &
\multicolumn{2}{c}{$0.00133_{-0.00018}^{+0.00013}$} \\
\multicolumn{2}{l}{Derived RV Parameters:} & APOGEE-2S & PFS\\
~~~Systemic Velocity\(^{d}\)\dotfill & $\gamma$ (m \unit{s^{-1}})\dotfill & 
$4863_{-86}^{+85}$ & $-7.73_{-6.93}^{+6.74}$\\
\sidehead{Derived Transit Parameters:}
~~~Time of Conjunction\dotfill & $T_C$ (BJD\textsubscript{TDB})\dotfill & \multicolumn{2}{c}{$2458326.279039_{-0.000984}^{+0.001024}$}\\
~~~Scaled Radius\dotfill & $R_{p}/R_{*}$ \dotfill & 
\multicolumn{2}{c}{$0.0912_{-0.0023}^{+0.0022}$}\\
~~~Scaled Semi-major Axis\dotfill & $a/R_{*}$ \dotfill &
\multicolumn{2}{c}{$8.08_{-0.22}^{+0.13}$}\\
~~~Orbital Inclination\dotfill & $i$ (degrees)\dotfill &
\multicolumn{2}{c}{$88.98_{-1.01}^{+0.70}$}\\
~~~Transit Impact Parameter\dotfill & $b$\dotfill &
\multicolumn{2}{c}{$0.14_{-0.10}^{+0.14}$}\\
~~~Transit Duration\dotfill & $T_{14}$ (hours)\dotfill &
\multicolumn{2}{c}{$5.99\pm0.07$}\\
\sidehead{Derived Planetary Parameters:}
~~~Mass\dotfill & $M_{p}$ (\unit{M_{J}})\dotfill &  \multicolumn{2}{c}{$1.75_{-0.17}^{+0.14}$}\\
~~~Radius\dotfill & $R_{p}$  (\unit{R_{J}}) \dotfill&
\multicolumn{2}{c}{$1.38\pm0.04$}\\
~~~Density\dotfill & $\rho_{p}$ (g \unit{cm^{-3}})\dotfill & \multicolumn{2}{c}{$0.83_{-0.11}^{+0.10}$}\\
~~~Surface Gravity\dotfill & $\log(g_{p})$ (cgs)\dotfill & \multicolumn{2}{c}{$3.215_{-0.045}^{+0.041}$}\\ 
~~~Equilibrium Temperature\(^{e}\)\dotfill & $T_{eq}$ (K)\dotfill & 
\multicolumn{2}{c}{$1493_{-32}^{+29}$}\\
\enddata
\tablenotetext{a}{{\tt SpecMatch-Emp} derived values.}
\tablenotetext{b}{{\tt EXOFASTv2} derived values using the \cite{Bailer-Jones2018} distance estimate as a prior.}
\tablenotetext{c}{Dilution is only considered for \tess{} light curves.}
\tablenotetext{d}{PFS provides relative velocities.}
\tablenotetext{e}{TOI-150.01 is assumed to be a blackbody.}
\normalsize
\end{deluxetable*}
\section{Discussion}\label{sec:discussion}
\begin{figure*}[!ht]
\epsscale{1.1}
\plotone{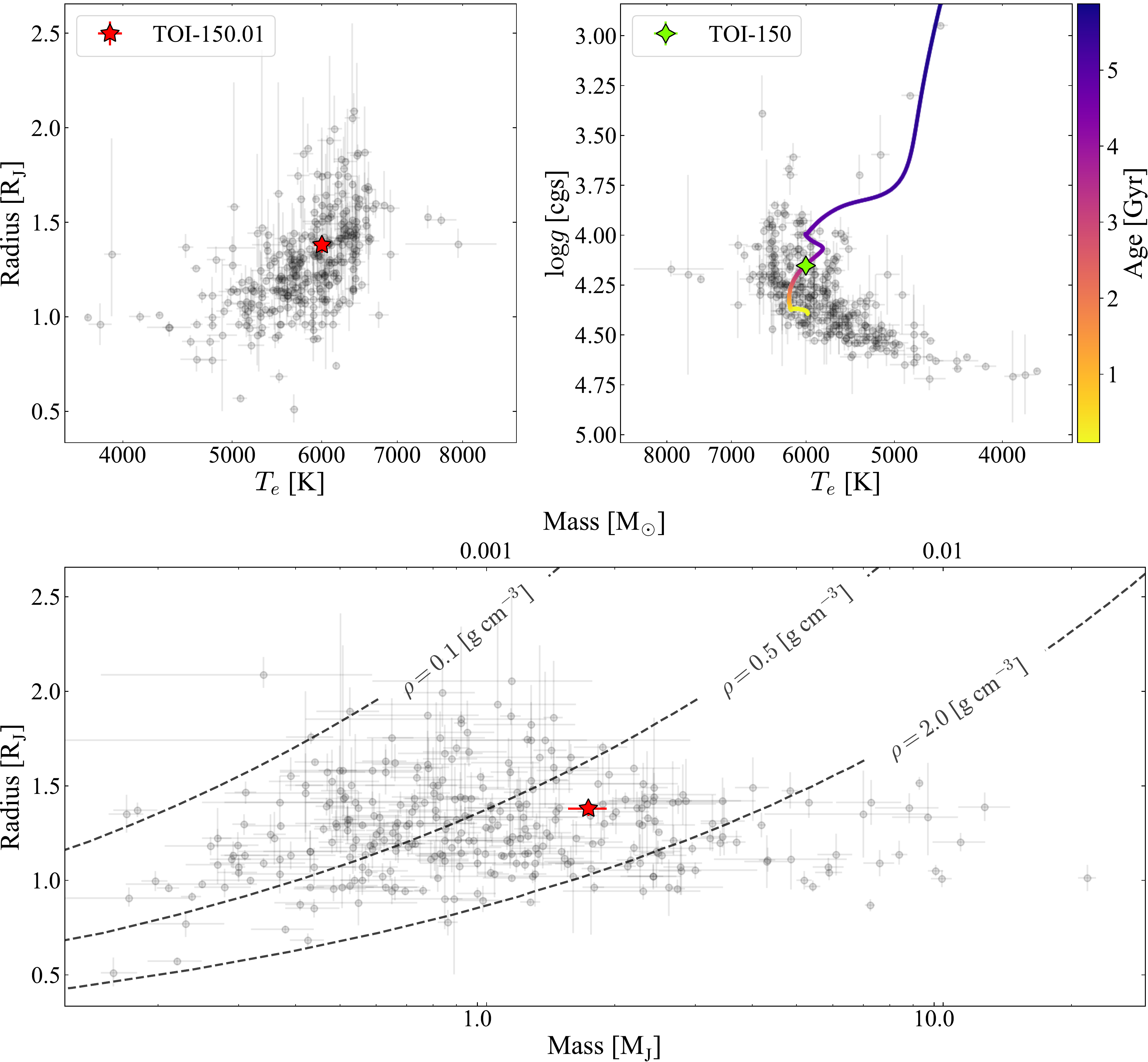}
\caption{TOI-150.01 compared to similar systems. The upper left panel compares TOI-150.01 to the distribution of radius and stellar effective temperature for gaseous exoplanets. The upper right panel places TOI-150, along with its best-matching MIST evolutionary track and other host stars, on a Hertzsprung-Russell diagram. The bottom panel shows TOI-150.01 on the mass-radius diagram for these planetary systems. The data were compiled from the \href{https://exoplanetarchive.ipac.caltech.edu/cgi-bin/TblView/nph-tblView?app=ExoTbls&config=planets}{NASA Exoplanet Archive} on 2019 May 2. \label{fig:f3}}
\end{figure*}

The modeling reveals that TOI-150 is in the second half of its core-hydrogen burning phase, hosting a transiting hot Jupiter in the \tess{} southern CVZ. While the current data cannot constrain the eccentricity of the orbit, it reveals the mass and radius are consistent with $1.38\pm0.04\unit{R_{J}}$ and $1.75_{-0.17}^{+0.14}\unit{M_{J}}$, respectively. Figure \ref{fig:f3} compares TOI-150 to known gas-giant exoplanets and their host stars. It is slightly inflated compared to exoplanets of comparable mass, which may be a result of stellar irradiation by the host star. 
\subsection{Opportunities for Further Characterization}
The APOGEE-2 and PFS RVs are consistent with the planetary nature of TOI-150.01 and constrain the orbital parameters. Subsequent high-precision spectroscopy will improve the precision of the orbital elements and the physical parameters of TOI-150.01. Additional spectra during transit would make it possible to constrain the apparent spin-orbit misalignment through analysis of the Rossiter-McLaughlin (RM) or reloaded RM effect \citep[e.g.,][]{Gaudi2007,Cegla2016}. Cross-correlation of the APOGEE-2S spectra with BT-Settl models reveals that TOI-150 is not a rapidly rotating star. When adopting a rotational velocity of \(v\sin i\sim2\unit{km\ s^{-1}}\), we estimate an RM effect amplitude of \(\sim16\unit{m\ s^{-1}}\). This is a value that can be detected using existing high-precision instruments.

TOI-150 will be observable for all 13 \tess{} Sectors in the southern ecliptic hemisphere. The long observational baseline will facilitate the search for additional planetary companions. While most known hot Jupiters have no detected close companions \citep[e.g.,][]{Dawson2018}, a full year of \tess{} photometry allows for a robust search of transiting companions and transit timing variations. The positive detection of a small, planetary companion to a hot Jupiter from the \kep{} mission \citep[e.g., Kepler-730;][]{Canas2019} required a long temporal baseline. Detecting a comparable planet will be very difficult for most \tess{} targets closer to the ecliptic with a \(\sim27\) day observational baseline. 

The overlap between the \tess{} and \textit{James Webb Space Telescope} (\jwst{}) CVZs also ensures TOI-150 will be observable for a large portion of each \jwst{} cycle. The long \tess{} observational baseline will establish a precise ephemeris that is necessary for potential emission spectroscopy using \jwst{}.
\subsection{Implications of Large-scale Spectroscopic Surveys for \tess{}}
A promising aspect of the \tess{} mission is that it will provide light curves for most of the sky that overlaps with existing, large-scale spectroscopic surveys, such as APOGEE-2. \cite{Troup2016} published 57 planetary candidates from previous APOGEE data and estimated the end of APOGEE-2 will have \(\sim1300\) substellar candidates orbiting different stellar populations and galactic environments. APOGEE-2 already has spectra for over 300,000 stars, most with \(\ge3\) observations, and has one northern and two southern programs observing the \tess{} CVZs. While the APOGEE-2 north and south spectrographs were designed for chemodynamics studies of the Milky Way, the precision is sufficient to detect giant exoplanets around bright stars \citep[e.g., HD 114762b, as shown in][]{Troup2016}. The two APOGEE-2 observations of TOI-150 represent a \(2\sigma\) detection of the hot Jupiter RV signal and were sufficient to exclude an eclipsing binary scenario and trigger subsequent PFS observations.

Once \tess{} begins observations in the northern ecliptic hemisphere, where there is more overlap with the bulk of APOGEE-2 data, it will be possible to effectively ``precover'' a \tess{}-detection for a hot Jupiter in APOGEE-2 velocimetry. While high-precision spectroscopy is still required to derive the most precise planetary mass, APOGEE-2 data can serve to (i) vet \tess{} candidates for eclipsing binaries \citep[e.g.,][]{Fleming2015}, (ii) provide spectroscopic parameters of the host star via ASPCAP \citep[e.g.,][]{Wilson2018}, and (iii) derive masses for giant planets when sufficient APOGEE observations are available. When \tess{} begins northern observations in late 2019, the planned 16th data release of SDSS (DR16) will provide a large, complementary data set for validating, and in some cases confirming, exoplanets around bright stars.
\section{Acknowledgements}
We thank the anonymous referee for a thoughtful reading of the manuscript and comments that improved the quality of this publication. C.I.C. and G.S. acknowledge support by NASA Headquarters under the NASA Earth and Space Science Fellowship Program through grants 80NSSC18K1114 and NNX16AO28H, respectively. C.I.C., C.F.B., and S.M. acknowledge support from NSF awards AST 1517592, 100667, 1126413, 1310885, N.D. acknowledges support from NFS award AST 1616684, and K.G.S acknowledges support from NASA XRP grant 17-XRP17 2-0024. S.W. thanks the Heising-Simons Foundation for their generous support.
Support for this work was provided by NASA through Hubble Fellowship grants HST-HF2-51399.001 (awarded to J.K.T.) and \#51386.01 (awarded to R.L.B.) by STScI, which is operated by the Association of Universities for Research in Astronomy, Inc., under contract NAS5-26555.
D.A.G.H. acknowledges support from the State Research Agency (AEI) of the Spanish Ministry of Science, Innovation and Universities (MCIU) and the European Regional Development Fund (FEDER) under grant AYA-2017-88254-P.
The research leading to these results has (partially) received funding from the European Research Council (ERC) under the EU’s Horizon 2020 research and innovation programme (grant agreement N$^\circ$670519: MAMSIE) and from the Fonds Wetenschappelijk Onderzoek - Vlaanderen under the grant agreement G0H5416N (ERC Opvangproject).

Part of this research was conducted using the Advanced CyberInfrastructure computational resources provided by The Institute for CyberScience at The Pennsylvania State University (\url{http://ics.psu.edu}), including the CyberLAMP cluster supported by NSF grant MRI-1626251.

Some of the data presented in this Letter were obtained from MAST. Support for MAST for non-HST data is provided by the NASA Office of Space Science via grant NNX09AF08G and by other grants and contracts. 2MASS is a joint project of the University of Massachusetts and IPAC at Caltech, funded by NASA and the NSF.
This Letter includes data collected by the \tess{} mission, which are publicly available from MAST. Funding for the \tess{} mission is provided by NASA's Science Mission directorate.
This work has made use of data from the European Space Agency (ESA) mission {\it Gaia} (\url{https://www.cosmos.esa.int/gaia}), processed by the {\it Gaia} Data Processing and Analysis Consortium (DPAC, \url{https://www.cosmos.esa.int/web/gaia/dpac/consortium}). Funding for the DPAC has been provided by national institutions, in particular the institutions participating in the {\it Gaia} Multilateral Agreement.

Funding for SDSS-IV has been provided by the Alfred P. Sloan Foundation, the U.S. Department of Energy Office of Science, and the Participating Institutions. SDSS-IV acknowledges support and resources from the Center for High-Performance Computing at the University of Utah. The SDSS website is \url{www.sdss.org}. SDSS-IV is managed by the Astrophysical Research Consortium for the Participating Institutions of the SDSS Collaboration including the Brazilian Participation Group, the Carnegie Institution for Science, Carnegie Mellon University, the Chilean Participation Group, the French Participation Group, Harvard-Smithsonian Center for Astrophysics, Instituto de Astrof\'isica de Canarias, The Johns Hopkins University, Kavli Institute for the Physics and Mathematics of the Universe (IPMU) / University of Tokyo, Lawrence Berkeley National Laboratory, Leibniz Institut f\"ur Astrophysik Potsdam (AIP), Max-Planck-Institut f\"ur Astronomie (MPIA Heidelberg), Max-Planck-Institut f\"ur Astrophysik (MPA Garching), Max-Planck-Institut f\"ur Extraterrestrische Physik (MPE), National Astronomical Observatories of China, New Mexico State University, New York University, University of Notre Dame, Observat\'ario Nacional / MCTI, The Ohio State University, Pennsylvania State University, Shanghai Astronomical Observatory, United Kingdom Participation Group, Universidad Nacional Aut\'onoma de M\'exico, University of Arizona, University of Colorado Boulder, University of Oxford, University of Portsmouth, University of Utah, University of Virginia, University of Washington, University of Wisconsin, Vanderbilt University, and Yale University.

\facilities{Du Pont (APOGEE-2S), \gaia{}, Magellan:Clay (PFS), \tess{}} 
\software{ AstroImageJ, 
{\tt astropy},
{\tt batman}, 
{\tt EXOFASTv2}, 
{\tt juliet}, 
{\tt MultiNest}, 
{\tt radvel}, 
{\tt SpecMatch-Emp}
}

%Track changes
\listofchanges

\end{document}